\documentclass[10pt,letterpaper]{article}

\usepackage[letterpaper,left=0.75in,right=0.75in,top=0.75in,bottom=0.85in,
            columnsep=0.25in,nohead]{geometry}
\usepackage{fix-cm}
\usepackage[T1]{fontenc}
\usepackage{newtxtext,newtxmath,courier}
\usepackage{helvet}
\usepackage[tracking=true,letterspace=16]{microtype}
\usepackage[table]{xcolor}
\usepackage{graphicx}

\usepackage{amsmath,amssymb,bm}
\usepackage{siunitx}
\usepackage{caption}
\usepackage{cite}

\definecolor{urlblue}{RGB}{46,46,177}
\usepackage[colorlinks=true,urlcolor=urlblue,linkcolor=black,citecolor=black,
            breaklinks=true,bookmarks=false]{hyperref}

\newcommand{\thpmt}{\ensuremath{\theta_{\mathrm{PMT}}}}
\newcommand{\dndl}{\mathrm{d}n/\mathrm{d}\lambda}
\newcommand{\eeff}{\ensuremath{\epsilon_{\mathrm{det}}}}

\renewcommand\refname{References}

\makeatletter
\renewcommand\@seccntformat[1]{\csname the#1\endcsname.\quad}

\renewcommand\section{\@startsection{section}{1}{0pt}%
  {-2.5ex plus -1ex minus -.2ex}{1.3ex plus .2ex}{\normalsize\bfseries\sffamily}}
\renewcommand\subsection{\@startsection{subsection}{2}{0pt}%
  {-2.25ex plus -1ex minus -.2ex}{1ex plus .2ex}{\normalsize\itshape\sffamily}}
\renewcommand\subsubsection{\@startsection{subsubsection}{3}{0pt}%
  {-2.25ex plus -1ex minus -.2ex}{1ex plus .2ex}{\small\upshape\sffamily}}
\makeatother

\newenvironment{backmatter}{\small%
  \newcommand{\bmsection}[1]{\par\medskip\noindent{\bfseries ##1.\enspace}}}{}

\makeatletter
\renewenvironment{thebibliography}[1]
  {\section*{\refname}%
   \list{\@arabic\c@enumiv.\hfill}%
     {\settowidth\labelwidth{\footnotesize 99.}%
      \setlength{\parsep}{-0.05in}%
      \setlength{\labelsep}{.25em}%
      \setlength{\itemsep}{0.05in}%
      \setlength{\listparindent}{0in}%
      \setlength{\itemindent}{0in}%
      \setlength{\leftmargin}{\labelwidth}%
      \addtolength{\leftmargin}{\labelsep}%
      \setlength{\rightmargin}{0in}%
      \usecounter{enumiv}%
      \let\p@enumiv\@empty
      \renewcommand\theenumiv{\@arabic\c@enumiv}}%
   \sloppy\clubpenalty4000\@clubpenalty\clubpenalty\widowpenalty4000%
   \sfcode`\.\@m\footnotesize}
  {\def\@noitemerr{\@latex@warning{Empty `thebibliography' environment}}%
   \endlist}
\makeatother

\begin{document}

\pagestyle{plain}

\twocolumn[{%
  {\fontsize{16pt}{18.5pt}\sffamily\bfseries\raggedright
   \textls{Generation and spectroscopic characterization of continuous-wave
   148.4-nm light in strontium tetraborate}\par\vskip.15in}%
  {\raggedright\sffamily\bfseries\scshape\large\boldmath
   Ming Guan,$^{1,5,\dagger}$ Yang Wang,$^{1,4,\dagger}$ Heng Tu,$^{2,4,\dagger}$
   Zhengrong Xiao,$^{1,4}$ Zhong Zuo,$^{1}$ Hengzhi Zhang,$^{1,4}$
   Kaifeng Zhou,$^{1,4}$ Songpo Xu,$^{1}$ Shaogang Yu,$^{1,4}$ Xin Tong,$^{1,4}$
   Jiyong Yao,$^{2,4}$ Jun Shen,$^{2,4}$ Zhiqiang Zhang,$^{3}$
   Linqiang Hua,$^{1,4,6}$ and Xiaojun Liu$^{1,4,7}$\vskip1ex\par}%
  {\raggedright\small\itshape
   $^{1}$Wuhan Institute of Physics and Mathematics, Innovation Academy for
   Precision Measurement Science and Technology, Chinese Academy of Sciences,
   Wuhan 430071, China\\
   $^{2}$Technical Institute of Physics and Chemistry, Chinese Academy of
   Sciences, Beijing 100190, China\\
   $^{3}$Hefei National Laboratory, University of Science and Technology of
   China, Hefei 230088, China\\
   $^{4}$University of Chinese Academy of Sciences, Beijing 100049, China\\
   $^{\dagger}$These authors contributed equally to this work.\par}%
  {\raggedright\footnotesize\itshape\color{urlblue}%
   $^{5}$guanming@apm.ac.cn\\
   $^{6}$hualq@apm.ac.cn\\
   $^{7}$xjliu@apm.ac.cn\par}%
  \vskip6pt
}]

\vskip1pc\noindent\textbf{\textsf{Abstract:}\space}
Continuous-wave (CW) radiation at 148.4~nm is required to drive the $^{229}$Th
isomeric transition of a solid-state nuclear clock, and frequency doubling of
296.8-nm light is its most direct source. We generate CW 148.4-nm light by random
quasi-phase matching in strontium tetraborate and characterize it with a compact
CaF$_2$ prism spectrometer, which reaches picowatt sensitivity without filters or
a grating and reads the wavelength from the rotation angle of a solar-blind
photomultiplier. The observed harmonic profile has an angular FWHM of
$0.36^{\circ}$, corresponding to a convolution-limited spectral width of about
\qty{0.6}{\nano\metre} rather than the intrinsic linewidth. The
directly estimated output is \qty{1.9}{\nano\watt} at room temperature; scaling
the measured temperature dependence gives an inferred \qty{4.8}{\nano\watt} at
the best observed temperature. The conversion varies by
an order of magnitude across the crystal and is strongly structured in
temperature: between \qty{30}{\degreeCelsius} and \qty{50}{\degreeCelsius} it
shows several maxima only a few degrees apart, so a small excursion moves the
output from a maximum to a minimum. These measurements provide valuable information on the realization and
characterization of a high-quality 148.4-nm light source, and pave the way for
the future development of high-performance nuclear optical clocks.

\vskip1pc

\section{Introduction}

The $^{229}$Th nucleus has a low-lying isomeric state at 8.4~eV, that is, at
148.4~nm, and it is currently the only nuclear transition within reach of table-top laser
technology \cite{peik2003}. Its energy is largely decoupled from the electronic
shell, so the transition should be exceptionally narrow, and the isomer could
serve as a solid-state clock, embedded in a vacuum-ultraviolet (VUV) transparent
crystal \cite{kazakov2012}, or as an ion clock that inherits the stability of an
optical atomic clock \cite{ludlow2015} while being far less sensitive to external
perturbations \cite{campbell2012}. The isomer has been detected through its
internal conversion decay \cite{vonderwense2016}, its excitation energy measured
to sub-eV precision \cite{seiferle2019}, and the transition pumped by X-rays
\cite{masuda2019}. Its radiative decay has been observed \cite{kraemer2023}, the
transition driven by a laser \cite{tiedau2024} and characterized
spectroscopically \cite{thielking2018,yamaguchi2024laser,shigekawa2026lifetime},
and its frequency ratio to a $^{87}$Sr clock measured with a VUV frequency comb
\cite{zhang2024frequency}. After the laser M\"ossbauer spectroscopy in a solid-state host
\cite{hiraki2026laser}, nuclear clock operation with 148-nm light from four-wave
mixing \cite{xiao2026,huang2026nuclear} or from solid-state VUV sources
\cite{toscani2026thorium,lal2025continuous,wang2026solid,wang2026continuous}
has also been reported. All of these experiments need narrow-line
continuous-wave (CW) radiation at 148.4~nm whose power, spectral purity and
stability can be measured and controlled \cite{vonderwense2019}.

The most direct route to such radiation is second-harmonic generation (SHG) of
296.8-nm light. Strontium tetraborate (SrB$_4$O$_7$, SBO) cannot be birefringently
phase-matched in this range, but the random orientation of its antiparallel
ferroelectric domains provides random quasi-phase matching (RQPM)
\cite{aleksandrovsky2010,baudrier2004}, which has already delivered coherent VUV
radiation down to 121~nm \cite{trabs2016}. Single-pass CW conversion in SBO
reaches 1.3~nW at 325~mW of pump power \cite{lal2025continuous}, 1.7~nW at
580~mW \cite{wu2026} and 8.1~nW at 920~mW \cite{wang2026continuous}.
Quasi-phase-matched BaMgF$_4$ reaches tens of picowatts at about \qty{1}{\watt}
\cite{Subramanian:26,bu2026}, while resonant conversion in Cd vapour has delivered
CW 148.4-nm radiation at the \qty{10}{\micro\watt} level with a linewidth of about
\qty{100}{\hertz} \cite{xiao2026,xiao2026proposal}. For a solid-state nuclear
clock, the laser source has to operate over timescales of hours. Its temperature
dependence, spatial uniformity and behaviour under prolonged irradiation therefore
affect the design of an experiment as directly as its output power; all three have
to be characterized before such a source can be used in a clock.

Such a characterization is not simply a matter of pointing a detector at the beam.
Molecular oxygen absorbs 148.4-nm light so strongly
\cite{watanabe1953} that a few millimetres of air already remove most of it, so
the light has to travel in nitrogen, helium or vacuum. Due to low conversion efficiency of RQPM ($\sim10^{-9}$), and the low efficiency of the VUV optical devices, the signal usually reduced to picowatt level on the detector, thus increasing the difficulty to detect the signal. Therefore, seldom spectrometers work in this wavelength domain.
 Reported solutions place a solar-blind CsI photomultiplier tube
(PMT) behind a stack of VUV bandpass filters in an evacuated chamber
\cite{wu2026}, or use a commercial vacuum grating spectrometer
\cite{lal2025continuous}. Both work well, but the filters fix the passband and cannot read the exact wavelength, while a grating instrument is limited by the strict geometric configuration, increasing the difficulties to detect the signal in the vacuum system.
 Thus, for the RQPM conversion of SBO in the VUV region, an setup which can tell the wavelength and the power is limited.

Here we report CW 148.4-nm generation in SBO and a complete spectroscopic
characterization of the light that it delivers, obtained with a compact prism
spectrometer built for the purpose. Instead of rejecting the fundamental with
coatings, we disperse the light with a CaF$_2$ prism and scan the spectrum with a
\qty{100}{\micro\metre} slit in front of a solar-blind PMT on a rotation stage, so
that the wavelength is encoded in the PMT angle; dichroic mirrors and a baffle
tube suppress the scattered pump before dispersion, the path sits in a chamber
that is evacuated and then filled with nitrogen, and the angle-to-wavelength
relation is calibrated. The setup can capture the signal photons, and detect the picowatt VUV siganl.
With this spectrometer, our measurements show that the harmonic follows the quadratic law expected for SHG, establish an absolute power, and measure how
the conversion depends on the crystal position, on the temperature, and on the
irradiation history. The most striking of these is the temperature response, which
is not a single smooth optimum but a sequence of maxima and minima only a few
degrees apart.

\begin{figure}[t]
  \centering
  \includegraphics[width=\linewidth]{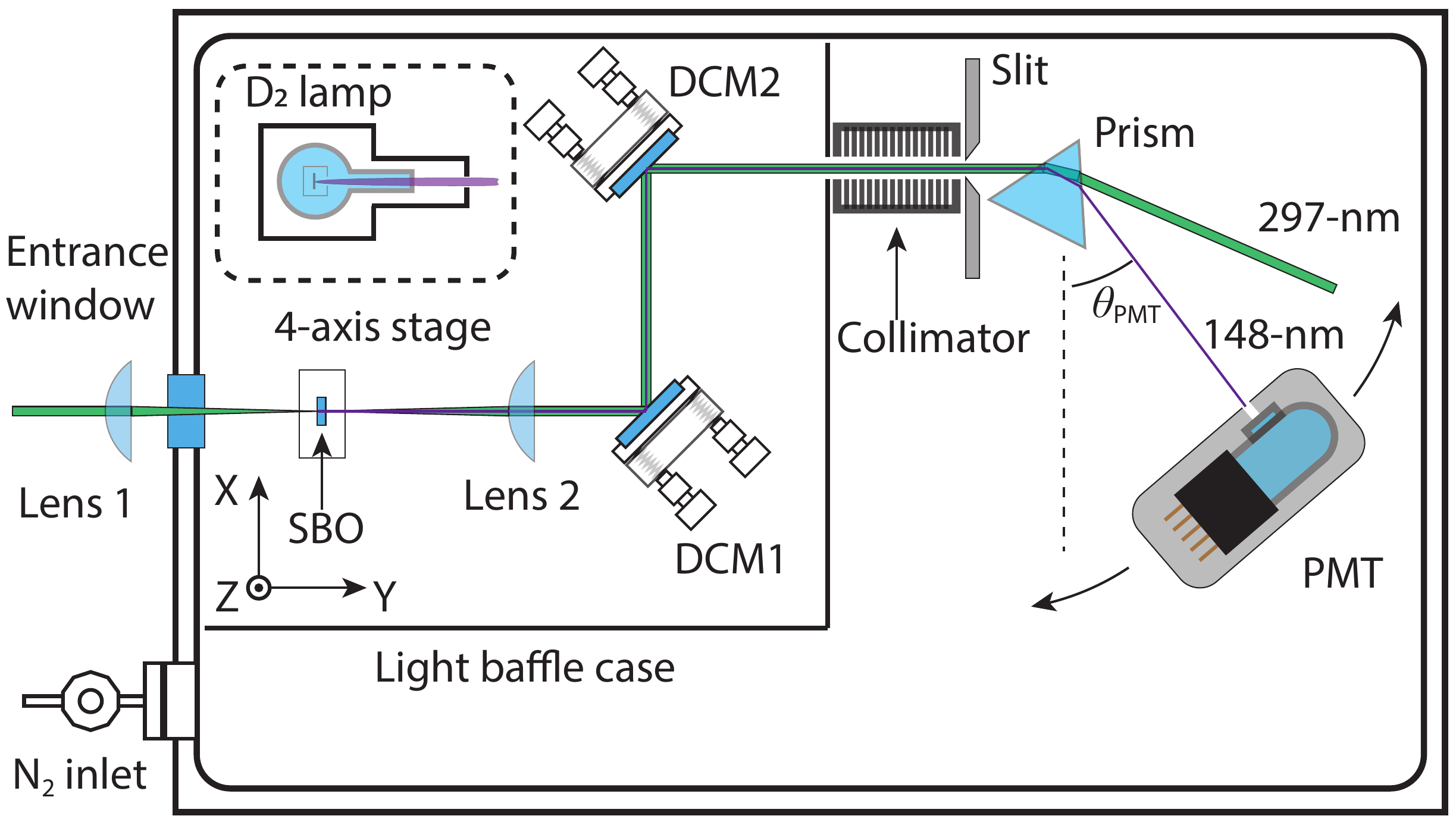}
  \caption{Experimental arrangement. The 296.8-nm pump is focused by Lens~1
    into the SBO crystal on a four-axis stage and the second harmonic is
    collimated by Lens~2; a pair of dichroic mirrors (DCM) folds the beam and
    rejects the residual pump, an \qty{80}{\milli\metre} baffle tube with a
    \qty{1}{\milli\metre} entrance slit limits the angular divergence (the
    second slit drawn behind it is fully open), the CaF$_2$ prism disperses the
    beam, and a solar-blind PMT behind a \qty{100}{\micro\metre} slit on a
    rotation stage records the intensity at the angle \thpmt. The deuterium
    lamp used for the calibration is drawn dashed. The optics sits in a chamber
    that is evacuated and then filled with nitrogen through the inlet, inside a
    light-baffle case.}
  \label{fig:setup}
\end{figure}

\section{Method}

\subsection{Crystal growth and sample preparation}

The SBO crystal was grown by the Czochralski method \cite{uecker2014historical} at
the Technical Institute of Physics and Chemistry, Chinese Academy of Sciences
(Beijing). High-purity (99.99\%) boric acid (H$_3$BO$_3$) and strontium carbonate
(SrCO$_3$) were weighed in stoichiometric ratio and mixed, and the furnace was
held close to the melting point of SrB$_4$O$_7$ while the crystal was pulled
along the $a$ axis at 2~mm/day. Ten days of growth gave a transparent boule of
$45\times10\times15$~mm$^3$. A \qty{1}{\milli\metre} thick $a$-cut plate was cut
from it, with only its surfaces polished and no coating applied; X-ray diffraction
confirmed the orientation before the SHG measurements.

\subsection{Optical layout}

The layout is shown in Fig.~\ref{fig:setup}. The 296.8-nm pump (Preci-Laser) is focused by Lens~1 ($f=300$\,mm) through the entrance window into the SBO
crystal, which sits on a four-axis stage (translation along XYZ, rotation around X)
that sets both the pump focus and the crystal region sampled by the beam. The
crystal and Lens~2 ($f=150$\,mm) are enclosed in a light-baffle case that
blocks scattered light originating from the pump.

Lens~2 collimates the output, and a pair of dichroic mirrors (DCM) folds the beam
into the dispersive arm; each reflection reflects the VUV and transmits part of
the pump out of the path, so the double bounce attenuates the residual fundamental
by about six orders of magnitude before the beam reaches the prism. An
\qty{80}{\milli\metre} baffle tube (the collimator in Fig.~\ref{fig:setup}) then
accepts a cone of about $0.7^{\circ}$ through its \qty{1}{\milli\metre} entrance
slit and keeps light scattered at larger angles away from the prism; the second
slit drawn behind it in Fig.~\ref{fig:setup} is fully open (\qty{2}{\milli\metre})
and does not act on the beam. The aperture that defines the instrument is the
\qty{100}{\micro\metre} slit directly in front of the PMT, which selects a small
angular and spatial fraction of the dispersed beam and therefore sets the sampling
of the spectrum.

The collimated beam enters a CaF$_2$ prism (PS863, Thorlabs) of apex angle
$A=60^{\circ}$. Refraction at the two faces obeys
\begin{equation}
  \begin{aligned}
    \sin i_1 &= n(\lambda)\sin r_1, &
    n(\lambda)\sin r_2 &= \sin i_2,\\
    A &= r_1 + r_2, &
    \delta &= i_1 + i_2 - A,
  \end{aligned}
  \label{eq:prism}
\end{equation}
where $n(\lambda)$ is the refractive index and $\delta$ the deflection angle.
Because $n$ rises steeply towards short wavelengths in the VUV, the 148.4-nm
beam is deviated more strongly than the 296.8-nm pump. A solar-blind CsI
PMT (R10454, Hamamatsu) is carried by a rotation stage (Lab65.RS, Multi-Field)
mounted below the prism, with its axis through the refracting corner; a rigid arm
links the PMT to the stage and points at the axis, so rotating the stage carries
the PMT around the prism at a fixed distance. The prism-to-PMT distance is about
\qty{150}{\milli\metre}, so one degree of \thpmt\ corresponds to
\qty{2.6}{\milli\metre} of travel. The PMT records the light as a function of the
angle \thpmt\ measured from the X direction (Fig.~\ref{fig:setup}), so that
$\delta = 90^\circ - \thpmt$.

The whole optical path is enclosed in a sealed chamber (Fig.~\ref{fig:setup}).
Before each run the chamber is pumped to about \qty{1e-3}{\pascal} and then
back-filled with dry nitrogen through the inlet, so the spectrometer can be run
either in vacuum or in a nitrogen atmosphere.

\subsection{Dispersion and wavelength calibration}
\label{sec:calib}

For a prism operated at minimum deviation, Eq.~(\ref{eq:prism}) gives the angular
dispersion as
\begin{equation}
  \left|\frac{\mathrm{d}\delta}{\mathrm{d}\lambda}\right| =
  \frac{2\sin(A/2)}{\sqrt{1-n^{2}(\lambda)\sin^{2}(A/2)}}\,
  \left|\frac{\mathrm{d}n}{\mathrm{d}\lambda}\right| ,
  \label{eq:dispersion}
\end{equation}
so the separation of two lines grows with the material dispersion $\dndl$ and
with the apex angle. CaF$_2$ is used because it combines transparency down to
about \qty{120}{\nano\metre} with a dispersion that is large in the VUV:
Eq.~(\ref{eq:dispersion}) gives about \qty{1.7}{\nano\metre/\degree} at
148.4~nm and \qty{30}{\nano\metre/\degree} at 296.8~nm.

\begin{figure}[t]
  \centering
  \includegraphics[width=\linewidth]{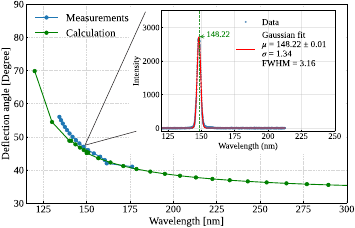}
  \caption{Wavelength calibration of the spectrometer. Deflection angles
    measured for the emission lines of a deuterium lamp (points) and
    calculated from Eq.~(\ref{eq:prism}) with the Sellmeier data of CaF$_2$
    (line); the deflection angle is $\delta = 90^\circ - \thpmt$. Inset: the
    \qty{148.22}{\nano\metre} feature of the dispersed lamp light, recorded with the reference
    spectrometer at the deflection angle around 46$^\circ$, with a Gaussian fit ($\mu = 148.22$~nm,
    $\sigma = \qty{1.34}{\nano\metre}$, FWHM $= \qty{3.2}{\nano\metre}$).}
  \label{fig:calib}
\end{figure}

The wavelength scale is calibrated with a deuterium lamp (L2D2, Hamamatsu) and a
reference VUV spectrometer (VS7550, Resonance Ltd.). The lamp is placed at the
position of DCM2 and its output is directed to the collimator, while the reference
spectrometer with a \qty{100}{\micro\metre} input aperture is mounted on the
rotation-stage arm; the inset of Fig.~\ref{fig:calib} shows the
\qty{148.22}{\nano\metre} feature, fitted with a Gaussian
($\mu = 148.22$~nm, $\sigma = \qty{1.34}{\nano\metre}$), recorded with the
reference spectrometer turned to the appropriate angle to the prism.
Figure~\ref{fig:calib} compares the deflection angles measured in this way between
130 and 185~nm (blue points) with the angles calculated from Eq.~(\ref{eq:prism})
using the Sellmeier data of CaF$_2$ (green points)
\cite{moyssides2000vuv,daimon2002}; between 140 and 180~nm the two agree to within
about $1.5^{\circ}$.
The calculation extends to 300~nm and thus covers the 296.8-nm pump. To map \thpmt\ to wavelength, each measured line centre is then
fitted as a function of \thpmt\ with a fourth-order polynomial,
\begin{equation}
  \lambda ( \thpmt)= \sum_{0}^{4}a_i \thpmt^i
  \label{eq:fit}
\end{equation}

with $a_0=3.651\times10^{3}$, $a_1=-3.627\times10^{2}$, $a_2=1.399\times10^{1}$, $a_3=-2.40\times10^{-1}$, and $a_4=1.552\times10^{-3}$, where \thpmt\ is in degrees
and $\lambda$ in nanometres; applied to the recorded spectra, the fit produces the
wavelength scale shown on their upper axes (Fig.~\ref{fig:scan}).

As a check at the two wavelengths of interest, the SHG beam itself provides an
internal reference: the 148.4-nm line measured below sits at
$\thpmt = 43.8^{\circ}$, which Eq.~(\ref{eq:fit}) maps to \qty{148.7}{\nano\metre},
and the 296.8-nm pump sits at $\thpmt = 55.0^{\circ}$, which the same calibration
maps to \qty{295}{\nano\metre}; both agree with the nominal wavelengths to within
the width of the lines.

\subsection{Detection efficiency and data acquisition}
\label{sec:efficiency}

Background is controlled at three stages: the light baffle blocks scattering at
the crystal, the dichroic pair removes most of the fundamental before dispersion,
and the angular selection of the PMT rejects light that is not deviated onto the
detector. The residual is measured with the PMT parked between the two lines;
with the pump on a bright RQPM spot, the 148.4-nm line reaches
$1.4\times10^{4}$~cps against $1.0\times10^{3}$~cps between the lines
(Fig.~\ref{fig:scan}).

The count rate is converted into a VUV power through the overall efficiency of
the detection channel,
\begin{equation}
  \eeff = \epsilon_{\mathrm{SBO}}\,
          \epsilon_{\mathrm{Lens2}}\,
          \epsilon_{\mathrm{DCM1}}\,
          \epsilon_{\mathrm{DCM2}}\,
          \epsilon_{\mathrm{prism}}\,
          Q_{\mathrm{PMT}} ,
  \label{eq:efficiency}
\end{equation}
where $\epsilon_{\mathrm{SBO}}\approx0.2$ is the transmission of the
\qty{148.4}{\nano\metre} light through the \qty{1}{\milli\metre} crystal,
$\epsilon_{\mathrm{Lens2}}=0.8$ and $\epsilon_{\mathrm{prism}}=0.8$ are the
transmissions of the collimating lens and of the two prism faces,
$\epsilon_{\mathrm{DCM1}}=\epsilon_{\mathrm{DCM2}}=0.6$ are the reflectivities
of the two dichroics at
148.4~nm, and $Q_{\mathrm{PMT}}\approx0.2$ is the quantum efficiency of the {CsI}
photocathode at this wavelength \cite{breskin1996}. The product is $\eeff\approx0.9\%$.

The \qty{100}{\micro\metre} slit in front of the PMT does not enter
Eq.~(\ref{eq:efficiency}). Because the detector is scanned across the line, the
rate integrated over the SHG profile counts every photon delivered to the PMT; the
slit only
lowers the rate recorded at any single angle.

The absolute SHG power then follows from the total number of photoelectrons
recorded across the line,
\begin{equation}
  P_{\mathrm{SHG}} = \frac{N\,E_{\gamma}}{\eeff},
  \label{eq:power}
\end{equation}
where $N$ is the integrated count rate, $E_{\gamma} = 8.4~\mathrm{eV} =
1.34\times10^{-18}~\mathrm{J}$ is the photon energy at 148.4~nm, and $\eeff$ is
the detection efficiency of Eq.~(\ref{eq:efficiency}).

The readout is a single-photon counting chain. The PMT output is amplified
($\times$25, SR445A, Stanford Research Systems) and digitized (PXIe-5160, National
Instruments); the waveforms are stored in binary, converted to ROOT files with
the CERN ROOT analysis package, and histogrammed to give the count rate as a
function of angle, that is, the spectrum.

\section{Results}
\label{sec:results}

\subsection{Spectral response and rejection of the pump}

\begin{figure}[t]
  \centering
  \includegraphics[width=\linewidth]{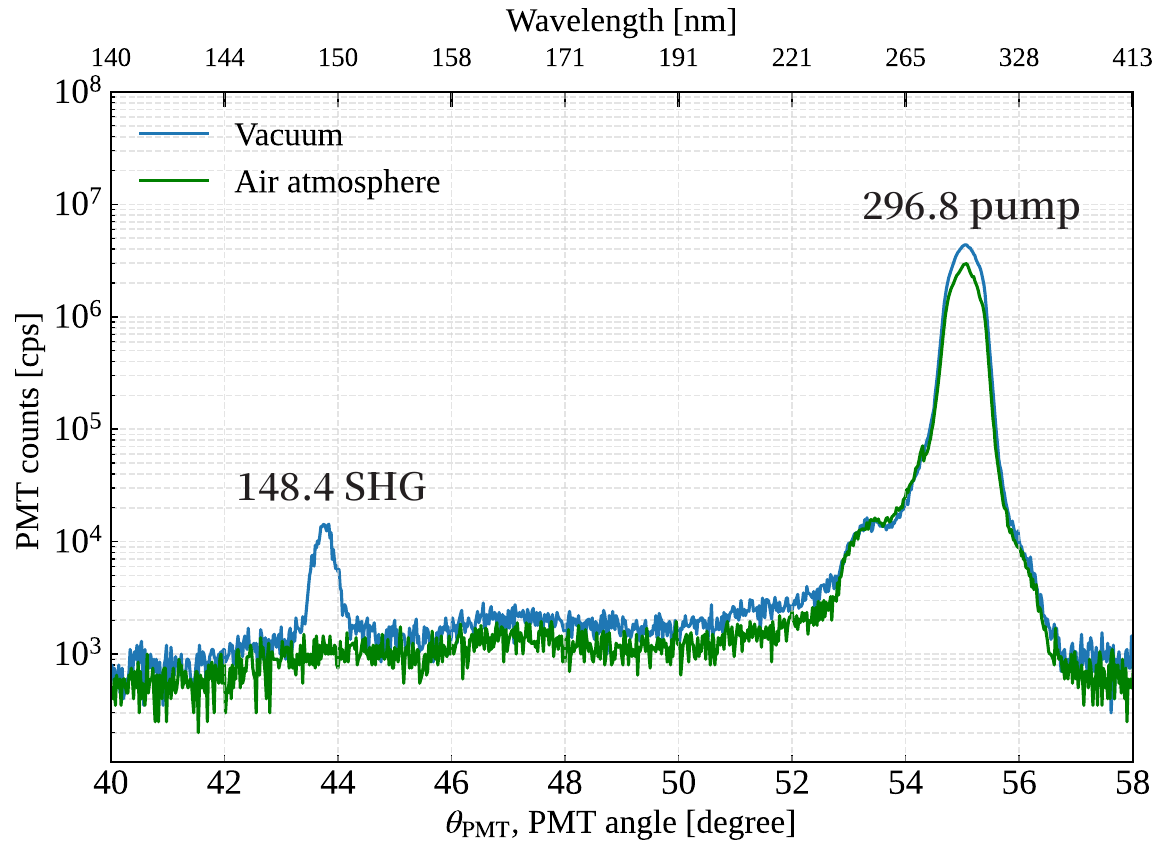}
  \caption{Angular scan of the spectrometer output in vacuum (blue) and with
    the chamber open to air (green). The pump line at 296.8~nm
    ($\thpmt \approx 55.0^\circ$) is about two orders of magnitude stronger
    than the second harmonic at 148.4~nm ($\thpmt \approx 43.8^\circ$); in air
    the harmonic is absorbed before it reaches the prism, while the pump is
    unaffected. The upper axis gives the wavelength obtained from the
    calibration of Sec.~\ref{sec:calib}.}
  \label{fig:scan}
\end{figure}

The complete spectral response of the instrument to the beam is shown in
Fig.~\ref{fig:scan}. The pump line appears at $\thpmt \approx 55.0^\circ$ and the
harmonic at $\thpmt \approx 43.8^\circ$, so that the two are separated by
$11.2^\circ$. The pump is also about two orders of magnitude brighter,
$1.5\times10^{6}~\mathrm{cps}$ against $1.4\times10^{4}~\mathrm{cps}$, with a
background of $1.0\times10^{3}$~cps between the lines; at the angle of the
harmonic the pump therefore contributes at most $7\%$ of the recorded rate.

Both lines have a Gaussian profile. The inset of Fig.~\ref{fig:quadratic} gives
$\sigma = 0.152^\circ$ for the harmonic, an angular FWHM of $0.36^\circ$, while
the pump line is about twice as wide at $0.7^\circ$. Divided by the local
dispersion of the calibration (upper axis of Fig.~\ref{fig:scan}), these widths
correspond to $\delta\lambda \approx \qty{0.6}{\nano\metre}$ at the harmonic and
\qty{20}{\nano\metre} at the pump. The first of the two is an upper bound rather
than a measure of the resolution: the \qty{100}{\micro\metre} slit subtends only
$0.038^\circ$ at the prism, an order of magnitude less than the recorded width, so
the profile is that of the beam size ($\phi \approx 1\mathrm{mm}$) and not of the
linewidth. For the purpose of this work the margin is what matters, and the
separation of the two lines, $11.2^\circ$, exceeds either width by an order of
magnitude.

The same figure contrasts a scan in vacuum (blue line) with one recorded with the chamber open
to air (green line). In air the 148.4-nm line disappears while the pump is unchanged, as
expected from the absorption length of molecular oxygen at this wavelength
\cite{watanabe1953}, which is a few tens of micrometres at atmospheric pressure;
the whole beam path, from the crystal to the PMT, therefore has to be purged or
evacuated. In vacuum, however, the SHG peak is in turn not stable: a first scan
across the line shows the peak, a second records less than half of it, and in a
third the peak is barely distinguishable from the background, although each scan
took only a few minutes. The rapid loss points to a radiation-induced change at
the irradiated spot, and prolonged irradiation in vacuum colours the whole crystal
[Fig.~\ref{fig:map}(c)]. Nitrogen is transparent at 148.4~nm and also helps to
carry heat away from the irradiated region, which protects the crystal. We
therefore acquired all spectra reported below with the chamber filled with dry
nitrogen.

\subsection{Quadratic scaling and line shape}

\begin{figure}[t]
  \centering
  \includegraphics[width=\linewidth]{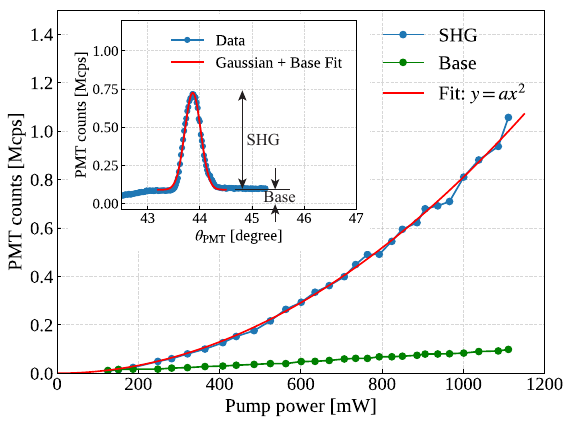}
  \caption{Quadratic dependence of the 148.4-nm signal on the 296.8-nm pump
    power. The second harmonic (blue) follows $y=ax^{2}$ over the whole range,
    while the background channel (green) grows linearly with the pump power.
    Inset: the 148.4-nm line recorded at the SHG angle, with a Gaussian fit
    ($\sigma = 0.152^\circ$, FWHM $= 0.36^\circ$); ``SHG'' and ``Base'' mark the
    amplitude of the Gaussian and the constant background, as used in the text.}
  \label{fig:quadratic}
\end{figure}

Figure~\ref{fig:quadratic} identifies the detected light as the second harmonic.
Its inset zooms in on \thpmt\ around the 148.4-nm line and fits the peak with a
Gaussian on a constant background; the amplitude is labelled ``SHG'' and the
constant ``Base'', standing for the 148.4-nm signal and for the background of
scattered pump light. Over the whole range explored, 100--1100~mW, the SHG count
rate (blue points) grows quadratically with the pump power, as a second-order
process should: a fit with $s = aP^{2}$ gives
$a = (695.2 \pm 9.9)\times10^{-9}$~Mcps\,mW$^{-2}$, where $s$ is the height of the
SHG line in megacounts per second (Mcps) and $P$ the 296.8-nm pump power in
milliwatts (mW). The background ``Base'' (green points) grows linearly with the
pump power, so the signal-to-background ratio improves steadily as the pump is
raised.

\subsection{Spatial uniformity and stability of the conversion}

\begin{figure}[t]
  \centering
  \includegraphics[width=\linewidth]{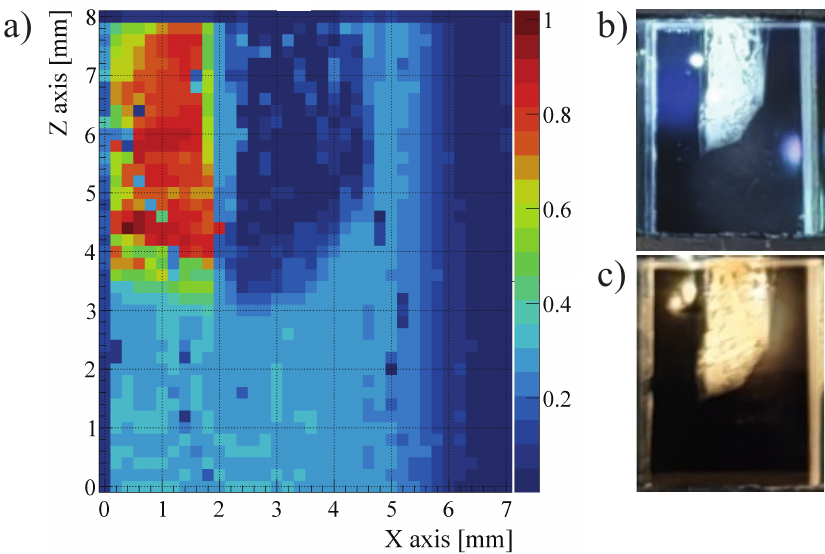}
  \caption{(a) Two-dimensional map of the normalized 148.4-nm count rate over
    the $X$--$Z$ plane of the SBO crystal, recorded with the pump focus fixed.
    (b),(c) Photographs of the same crystal before and after $\sim$30~h of
    irradiation at $\sim$1~W in vacuum; the yellow-brown colour in (c) is
    radiation-induced colour centres.}
  \label{fig:map}
\end{figure}

Since the 148.4-nm signal comes from RQPM, the conversion is far from uniform over
the crystal. Figure~\ref{fig:map}(a) maps the 148.4-nm count rate with the PMT
moved to the SHG maximum of the inset of Fig.~\ref{fig:quadratic} and the sample
translated in its $X$-$Z$ plane with the pump focus fixed (Fig.~\ref{fig:setup}).
The RQPM signal is concentrated in a region around $X\approx 1$~mm and
$Z\approx 4$--$8$~mm, where the recorded rate reaches the upper end of the colour
scale, and it falls to the background for $X>2$~mm and $Z<3.5$~mm; the contrast
between the two areas is an order of magnitude. Such a map is the spatial
fingerprint of RQPM \cite{aleksandrovsky2010,trabs2016}: the inhomogeneity sampled
by the beam is an average over randomly oriented domains, so translating the
crystal by a few millimetres samples a different realization of the same
ensemble. In practice the four-axis stage is not only an alignment tool but the
control that sets the VUV power, and Fig.~\ref{fig:map}(a) is the scan by which a
useful region is found.

Long-term operation changes the crystal itself. Figures~\ref{fig:map}(b)
and~\ref{fig:map}(c) show the same crystal before and after $\sim$30~h of
irradiation at $\sim$1~W in vacuum. The initially colourless crystal has turned
yellow-brown: colour centres absorbing in the visible have been created by the
pump radiation, consistent with the defect-related absorption of SBO studied in
Ref.~\cite{yang2026}. A CW source has to sit on the same crystal region for hours,
so this is the process that will eventually limit the VUV power.

\subsection{Temperature dependence of the conversion}

\begin{figure}[t]
  \centering
  \includegraphics[width=\linewidth]{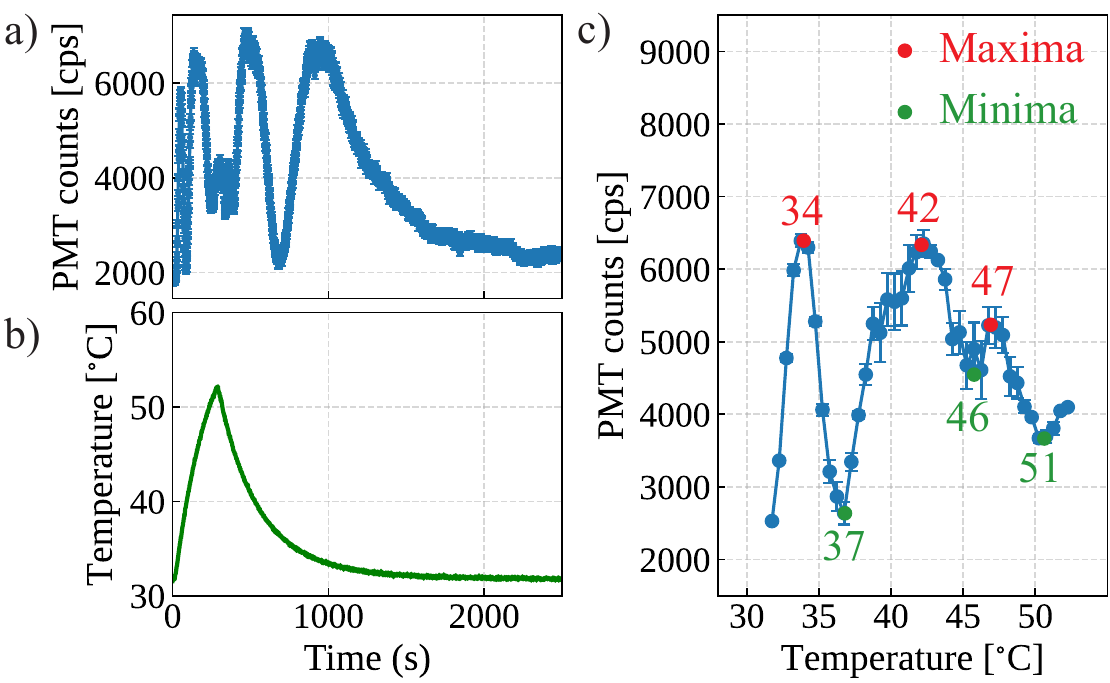}
  \caption{Thermal behaviour of the SHG signal, measured in nitrogen. (a) Count
    rate at the 148.4-nm line and (b) temperature of the SBO sample, recorded
    simultaneously as a function of time after the pump is switched on; (c) the
    count rate of (a) plotted against the temperature of (b). The rate passes
    through several local maxima (red) and minima (green) only a few degrees
    apart.}
  \label{fig:temp}
\end{figure}

The conversion efficiency is also sensitive to the crystal temperature.
Figure~\ref{fig:temp} shows the 148.4-nm count rate and the sample temperature,
recorded simultaneously in nitrogen as the crystal is heated to approximately
\qty{52}{\degreeCelsius} and subsequently allowed to cool. Over this temperature
excursion, the count rate varies from $2.5\times10^{3}$ to
$6.3\times10^{3}~\mathrm{cps}$, corresponding to a factor of 2.5. When plotted
against the sensor-read temperature [Fig.~\ref{fig:temp}(c)], reconstructed from
the two time traces, the data exhibit a hysteresis-like loop rather than a
single-valued dependence, indicating that the optical response is correlated with
the measured sample temperature and may also depend on the thermal history.

A second notable feature is the multi-peaked temperature dependence, which, to our
knowledge, has not previously been reported for second-harmonic generation in SBO.
Rather than exhibiting a single smooth maximum, the signal displays multiple peaks
and valleys. In the time trace [Fig.~\ref{fig:temp}(a)], the count rate passes
through at least three maxima before entering a slowly decaying tail. When plotted
against temperature [Fig.~\ref{fig:temp}(c)], the same data reveal local maxima
near \qty{34}{\degreeCelsius}, \qty{42}{\degreeCelsius} and
\qty{47}{\degreeCelsius}, separated by local minima near
\qty{37}{\degreeCelsius}, \qty{46}{\degreeCelsius} and
\qty{51}{\degreeCelsius}. The neighbouring extrema are separated by only a few
degrees, while the count-rate contrast between a maximum and an adjacent minimum
approaches the overall factor of 2.5 observed during the temperature excursion.
Notably, after the minimum near \qty{51}{\degreeCelsius} the count rate turns
upward again, which suggests additional structure beyond the range explored here.
Further measurements over an extended temperature range will be needed to
characterize these features and clarify their underlying mechanisms.

A single thermo-optic phase-matching condition cannot produce such a structure: it
would give one broad maximum, displaced smoothly by the temperature. A natural
reading is that the random domain structure samples a distribution of
quasi-phase-matching periods, so that different groups of domains satisfy the
matching condition at different temperatures; the crystal then behaves as a set of
partially independent converters that are switched on and off in succession as the
temperature is swept, and the recorded curve is their sum. The same picture,
together with the thermal inertia of the sample, the thermal expansion of the
random domain structure, a thermo-optic change of the RQPM condition and
pump-induced heating, also accounts for the lag of the optical response behind the
sensor reading in Fig.~\ref{fig:temp}(c). Whichever mechanism dominates, the
consequence for a future experiment is direct: the temperature has to be held on a
chosen maximum and not merely held constant.

\subsection{Absolute power and overall performance}

The absolute power follows from the integrated rate. The line recorded through
the \qty{100}{\micro\metre} slit is a Gaussian in \thpmt\ with
$\sigma = 0.15^{\circ}$, and its integral is about twelve times the peak of the
recorded histogram. At the highest pump power the peak reaches
$1.1~\mathrm{Mcps}$, so $13~\mathrm{Mcps}$ of 148.4-nm photons reach the PMT; with
the detection efficiency of Eq.~(\ref{eq:efficiency}), Eq.~(\ref{eq:power}) gives
\qty{1.9}{\nano\watt} at 148.4~nm generated in the crystal. Since the signal can
be raised by a factor of 2.5 by tuning the crystal to its optimum temperature
(Fig.~\ref{fig:temp}), the achievable power is about \qty{4.8}{\nano\watt}.

\section{Discussion}

The prism spectrometer trades spectral range for simplicity: it disperses the
whole 120--300~nm window, so one detector can follow any line in it without
changing filters or refocusing, at the price of a point-by-point scan. Because the
slit is an order of magnitude narrower than the image of the line, the resolution
is set by the divergence of the collimated beam, and a wider slit would raise the
count rate per point without loss of resolution; coating the prism faces, or
chopping the pump and detecting with a lock-in amplifier, would shorten the scan
and lower the background. The same arrangement applies to any VUV source that has
to be characterized before it is brought onto the nucleus.

We have demonstrated a CW 148.4-nm light source delivering an output power of
\qty{1.9}{\nano\watt} at room temperature, increasing to \qty{4.8}{\nano\watt}
when the crystal is maintained at its optimum temperature. This output power
places the source within the range relevant to absorption spectroscopy of the
${}^{229}\mathrm{Th}$ isomer in a host crystal
\cite{wang2026solid,lal2025continuous}, and the recorded line width establishes an
upper bound of $\delta\lambda \approx \qty{0.6}{\nano\metre}$ on its spectral
width.

Bringing this source to the nucleus will require holding the crystal on a chosen
maximum for hours while monitoring the conversion with the beam on the sample,
which the present arrangement already permits.  Three properties of the conversion have to be handled in any clock design.
First, the usable aperture is small: the conversion varies by an order of
magnitude over the crystal and is concentrated in a region about a
millimetre across, so the position of the crystal is a control parameter of the
source and not a fixed property. Second, the temperature dependence is
multi-peaked, with extrema a few degrees apart, so that stabilizing the source to a
single set point is not enough; the point has to sit on a chosen maximum and be
held there to a fraction of the peak spacing. Third, prolonged irradiation in
vacuum colours the crystal and reduces the output within hours, which limits the
operating life of a solid-state emitter and favours operation in nitrogen, at
reduced pump power, or with periodic annealing.

\section{Conclusion}

We have generated CW 148.4-nm light in strontium tetraborate by RQPM and
characterized it spectroscopically with a view to the $^{229}$Th nuclear
transition. A compact prism spectrometer built in this work reaches picowatt
sensitivity without filters or a grating; calibrated against a deuterium lamp, it
separates the harmonic from the 296.8-nm pump by angle alone and measures an
absolute power of \qty{1.9}{\nano\watt} at 148.4~nm under pumping by about
\qty{1}{\watt} at 296.8~nm. The harmonic profile has an angular FWHM of
$0.36^{\circ}$, corresponding to a convolution-limited spectral width of about
\qty{0.6}{\nano\metre} rather than the intrinsic linewidth; scaling
the measured temperature dependence gives \qty{4.8}{\nano\watt} at the best
observed temperature. The SHG conversion depends quadratically on the pump power
and is far from uniform: it varies by an order of
magnitude over the crystal, decays under prolonged irradiation in vacuum, and,
most importantly, passes through several local maxima and minima within a
\qty{20}{\degreeCelsius} wide temperature window, so that the source has to be
stabilized on a chosen maximum rather than merely held constant.
These measurements provide valuable information on the realization and
characterization of a high-quality 148.4-nm light source, and pave the way for
the future development of high-performance nuclear optical clocks.

\begin{backmatter}

\bmsection{Funding}
Strategic Priority Research Program of the Chinese Academy of Sciences (XDB0920000);
National Natural Science Foundation of China (12604402, 12404314, U21A20435);
CAS Project for Young Scientists in Basic Research (YSBR-055);
Postdoctoral Fellowship Program of CPSF (GZB20240765);
China Postdoctoral Science Foundation (2024M763347, 2025HBBSHXF017);
Hubei Provincial Natural Science Foundation of China (2025AFA004, 2026AFA088).

\bmsection{Acknowledgment}
We thank Prof. Yunquan Liu of Peking University for providing the VUV spectrometer for the wavelength calibration.

\bmsection{Disclosures}
The authors declare no conflicts of interest.

\bmsection{Data availability}
Data underlying the results presented in this paper are not publicly available
at this time but may be obtained from the authors upon reasonable request.

\end{backmatter}


\begin{thebibliography}{10}
\newcommand{\enquote}[1]{``#1''}

\bibitem{peik2003}
E.~Peik and C.~Tamm, \enquote{Nuclear laser spectroscopy of the 3.5 {eV} transition in {Th}-229,} \href{https://doi.org/10.1209/epl/i2003-00210-x}{Europhysics Letters \textbf{61}, 181--186} (2003).

\bibitem{kazakov2012}
G.~A. Kazakov, V.~V. Litvinov, V.~I. Romanenko \emph{et~al.}, \enquote{Performance of a {$^{229}$}thorium solid-state nuclear clock,} \href{https://doi.org/10.1088/1367-2630/14/8/083019}{New Journal of Physics \textbf{14}, 083019} (2012).

\bibitem{ludlow2015}
A.~D. Ludlow, M.~M. Boyd, J.~Ye \emph{et~al.}, \enquote{Optical atomic clocks,} \href{https://doi.org/10.1103/RevModPhys.87.637}{Reviews of Modern Physics \textbf{87}, 637--701} (2015).

\bibitem{campbell2012}
C.~J. Campbell, A.~G. Radnaev, A.~Kuzmich, \emph{et~al.}, \enquote{Single-ion nuclear clock for metrology at the 19th decimal place,} \href{https://doi.org/10.1103/PhysRevLett.108.120802}{Physical Review Letters \textbf{108}, 120802} (2012).

\bibitem{vonderwense2016}
L.~von~der Wense, B.~Seiferle, M.~Laatiaoui, \emph{et~al.}, \enquote{Direct detection of the {$^{229}$Th} nuclear clock transition,} \href{https://doi.org/10.1038/nature17669}{Nature \textbf{533}, 47--51} (2016).

\bibitem{seiferle2019}
B.~Seiferle, L.~von~der Wense, P.~V. Bilous \emph{et~al.}, \enquote{Energy of the {$^{229}$Th} nuclear clock transition,} \href{https://doi.org/10.1038/s41586-019-1533-4}{Nature \textbf{573}, 243--246} (2019).

\bibitem{masuda2019}
T.~Masuda, A.~Yoshimi, A.~Fujieda \emph{et~al.}, \enquote{{X}-ray pumping of the {$^{229}$Th} nuclear clock isomer,} \href{https://doi.org/10.1038/s41586-019-1542-3}{Nature \textbf{573}, 238--242} (2019).

\bibitem{kraemer2023}
S.~Kraemer, J.~Moens, M.~Athanasakis-Kaklamanakis \emph{et~al.}, \enquote{Observation of the radiative decay of the {$^{229}$Th} nuclear clock isomer,} \href{https://doi.org/10.1038/s41586-023-05894-z}{Nature \textbf{617}, 706--710} (2023).

\bibitem{tiedau2024}
J.~Tiedau, M.~V. Okhapkin, K.~Zhang \emph{et~al.}, \enquote{Laser excitation of the {$^{229}$Th} nucleus,} \href{https://doi.org/10.1103/PhysRevLett.132.182501}{Physical Review Letters \textbf{132}, 182501} (2024).

\bibitem{thielking2018}
J.~Thielking, M.~V. Okhapkin, D.~A. Gorbachev \emph{et~al.}, \enquote{Laser spectroscopic characterization of the nuclear-clock isomer {$^{229\mathrm{m}}$Th},} \href{https://doi.org/10.1038/s41586-018-0011-8}{Nature \textbf{556}, 321--325} (2018).

\bibitem{yamaguchi2024laser}
A.~Yamaguchi, Y.~Shigekawa, H.~Haba, \emph{et~al.}, \enquote{Laser spectroscopy of triply charged {$^{229}$Th} isomer for a nuclear clock,} \href{https://doi.org/10.1038/s41586-024-07296-1}{Nature \textbf{629}, 62--66} (2024).

\bibitem{shigekawa2026lifetime}
Y.~Shigekawa, A.~Yamaguchi, K.~Tokoi, \emph{et~al.}, \enquote{Lifetime of the singly charged {$^{229}$Th} nuclear isomer,} \href{https://doi.org/10.1038/s41567-026-03251-1}{Nature Physics \textbf{22}, 832--837} (2026).

\bibitem{zhang2024frequency}
C.~Zhang, T.~Ooi, J.~S. Higgins, \emph{et~al.}, \enquote{Frequency ratio of the $^{229m}${Th} nuclear isomeric transition and the $^{87}${Sr} atomic clock,} \href{https://doi.org/10.1038/s41586-024-07839-6}{Nature \textbf{633}, 63--70} (2024).

\bibitem{hiraki2026laser}
T.~Hiraki, T.~Masuda, S.~Takatori, \emph{et~al.}, \enquote{Laser {M}{\"o}ssbauer spectroscopy of {$^{229}$Th} in {CaF$_2$},} \href{https://doi.org/10.1126/science.aea7978}{Science \textbf{393}, 795--799} (2026).

\bibitem{xiao2026}
Q.~Xiao, G.~Penyazkov, X.~Li \emph{et~al.}, \enquote{Continuous-wave narrow-linewidth vacuum ultraviolet laser source,} \href{https://doi.org/10.1038/s41586-026-10107-4}{Nature \textbf{650}, 852--856} (2026).

\bibitem{huang2026nuclear}
B.~Huang, G.~Yan, Q.~Xiao, \emph{et~al.}, \enquote{A nuclear clock synchronized to {$^{229}$Th},} \href{https://doi.org/10.1038/s41586-026-11122-1}{Nature \textbf{658}, 903--908} (2026).

\bibitem{toscani2026thorium}
L.~Toscani De~Col, T.~Riebner, I.~Morawetz, \emph{et~al.}, \enquote{A thorium-229 optical nuclear clock with feedback loop,} \href{https://doi.org/10.1038/s41586-026-11084-4}{Nature \textbf{658}, 897--902} (2026).

\bibitem{lal2025continuous}
V.~Lal, M.~V. Okhapkin, J.~Tiedau, \emph{et~al.}, \enquote{Continuous-wave laser source at the 148 nm nuclear transition of {Th}-229,} \href{https://doi.org/10.1364/OPTICA.574489}{Optica \textbf{12}, 1971--1974} (2025).

\bibitem{wang2026solid}
P.~Wang, X.-F. Yin, H.~Wang, \emph{et~al.}, \enquote{A solid-state nuclear clock based on {VUV} absorption spectroscopy of {$^{229}$Th},} \href{https://doi.org/10.48550/arXiv.2610.10056}{arXiv preprint arXiv:2610.10056 } (2026).

\bibitem{wang2026continuous}
H.~Wang, Z.~Tan, C.~Huang, \emph{et~al.}, \enquote{Continuous-wave 148.4-nm generation in strontium tetraborate for nuclear absorption spectroscopy,} \href{https://doi.org/10.48550/arXiv.2610.18470}{arXiv preprint arXiv:2610.18470 } (2026).

\bibitem{vonderwense2019}
L.~von~der Wense and B.~Seiferle, \enquote{The 229-thorium isomer: doorway to the road from the atomic clock to the nuclear clock,} \href{https://doi.org/10.1088/1361-6455/ab29b8}{Journal of Physics B: Atomic, Molecular and Optical Physics \textbf{52}, 203001} (2019).

\bibitem{aleksandrovsky2010}
A.~S. Aleksandrovsky, A.~M. Vyunishev, A.~I. Zaitsev, and V.~V. Slabko, \enquote{Random quasi-phase-matched conversion of broadband radiation in a nonlinear photonic crystal,} \href{https://doi.org/10.1103/PhysRevA.82.055806}{Physical Review A \textbf{82}, 055806} (2010).

\bibitem{baudrier2004}
M.~Baudrier-Raybaut, R.~Ha{\"i}dar, P.~Kupecek, \emph{et~al.}, \enquote{Random quasi-phase-matching in bulk polycrystalline isotropic nonlinear materials,} \href{https://doi.org/10.1038/nature03027}{Nature \textbf{432}, 374--376} (2004).

\bibitem{trabs2016}
P.~Trabs, F.~Noack, A.~S. Aleksandrovsky, \emph{et~al.}, \enquote{Generation of coherent radiation in the vacuum ultraviolet using randomly quasi-phase-matched strontium tetraborate,} \href{https://doi.org/10.1364/OL.41.000618}{Optics Letters \textbf{41}, 618--621} (2016).

\bibitem{wu2026}
Y.~Wu, P.~Yang, L.~Yan, \emph{et~al.}, \enquote{Continuous-wave 148.4-nm generation in top-seeded solution-grown strontium tetraborate,} \href{https://doi.org/10.48550/arXiv.2610.06477}{arXiv:2610.06477} (2026).

\bibitem{Subramanian:26}
K.~Subramanian, H.~Tanaka, S.~J. Herr, \emph{et~al.}, \enquote{Generation of continuous-wave laser light at 148.4 nm using cavity-enhanced second-harmonic generation in {BaMgF$_4$},} \href{https://doi.org/10.1364/OPTICA.609757}{Optica \textbf{13}, 2089--2095} (2026).

\bibitem{bu2026}
Y.~Bu, X.~Wang, R.~Gu \emph{et~al.}, \enquote{Continuous-wave 148.4 nm generation in quasi-phase matched {BaMgF$_4$},} \href{https://doi.org/10.48550/arXiv.2610.18519}{arXiv:2610.18519} (2026).

\bibitem{xiao2026proposal}
Q.~Xiao, G.~Penyazkov, R.~Yu, \emph{et~al.}, \enquote{Proposal for the generation of continuous-wave vacuum-ultraviolet laser light for {Th}-229 isomer precision spectroscopy,} \href{https://doi.org/10.1103/cw5h-644b}{Physical Review Applied \textbf{25}, 024034} (2026).

\bibitem{watanabe1953}
K.~Watanabe, E.~C.~Y. Inn, and M.~Zelikoff, \enquote{Absorption coefficients of oxygen in the vacuum ultraviolet,} \href{https://doi.org/10.1063/1.1699104}{Journal of Chemical Physics \textbf{21}, 1026--1030} (1953).

\bibitem{uecker2014historical}
R.~Uecker, \enquote{The historical development of the {Czochralski} method,} \href{https://doi.org/10.1016/j.jcrysgro.2013.11.095}{Journal of Crystal Growth \textbf{401}, 7--24} (2014).

\bibitem{moyssides2000vuv}
P.~Moyssides, S.~Maltezos, and E.~Fokitis, \enquote{A {VUV} prism spectrometer for {RICH} radiator refractometry,} \href{https://doi.org/10.1080/09500340050080312}{Journal of Modern Optics \textbf{47}, 1693--1706} (2000).

\bibitem{daimon2002}
M.~Daimon and A.~Masumura, \enquote{High-accuracy measurements of the refractive index and its temperature coefficient of calcium fluoride in a wide wavelength range from 138 to 2326 nm,} \href{https://doi.org/10.1364/AO.41.005275}{Applied Optics \textbf{41}, 5275--5287} (2002).

\bibitem{breskin1996}
A.~Breskin, \enquote{{CsI} {UV} photocathodes: history and mystery,} \href{https://doi.org/10.1016/0168-9002(95)01145-5}{Nuclear Instruments and Methods in Physics Research A \textbf{371}, 116--136} (1996).

\bibitem{yang2026}
R.~Yang, G.~Deng, C.~Huang \emph{et~al.}, \enquote{Experimental and ab initio study on the origin of ultraviolet absorption in {SrB$_4$O$_7$} crystal,} \href{https://doi.org/10.1016/j.physb.2026.419035}{Physica B \textbf{740}, 419035} (2026).

\end{thebibliography}
\end{document}